\documentclass[useAMS,usegraphicx,usenatbib]{mn2e}

\usepackage{graphicx}
\usepackage{times}
\usepackage{amssymb}
\usepackage{amsmath}
\newif\ifAMStwofonts
\AMStwofontstrue

\def\pmn{{PMN~J0525--3343}}
\def\gb{{GB~B1428+4217}}
\def\rx{{RX~J1028.6--0844}}
\def\pks{{PKS~2126--0158}}
\def\asca{{\it ASCA}}
\def\xmm{{\it XMM-Newton}}
\def\xmmtitle{{\it XMM-NEWTON}}
\def\bepposax{{\it BeppoSAX}}
\def\rosat{{\it ROSAT}}

\def\ks{{\rm\thinspace ks}}

\def\a{{\rm\thinspace\AA}}

\def\pcmsq{\hbox{$\rm\thinspace cm^{-2}$}}

\def\ctsps{\hbox{$\rm\thinspace cts~s^{-1}$}}

\def\kev{{\rm\thinspace keV}}

\def\ergpcmsqps{\hbox{$\rm\thinspace erg~cm^{-2}~s^{-1}$}}

\def\ergcmps{\hbox{$\rm\thinspace erg~cm~s^{-1}$}}

\title{\xmm\ observations of \gb: confirmation of intrinsic soft X-ray absorption}
\author[M. A. Worsley et al.]
{\parbox[]{6.in}
{M.~A. Worsley$^{1}$\thanks{E-mail: maw@ast.cam.ac.uk}, A.~C. Fabian$^{1}$, A. Celotti$^{2}$ and K. Iwasawa$^{1}$}\\
\footnotesize
$^{1}$Institute of Astronomy, Madingley Road, Cambridge CB3 0HA\\
$^{2}$S.~I.~S.~S.~A., via Beirut 2-4, I-34014 Trieste, Italy\\
}

\begin{document}
\maketitle

\label{firstpage}

\begin{abstract}
We report the results of \xmm\ observations of the X-ray bright, radio-loud blazar \gb\ at a redshift of $z=4.72$. We confirm the presence of soft X-ray spectral flattening at energies $\lesssim0.7\kev$ as reported in previous \rosat\ and \bepposax\ observations. At hard X-ray energies the spectrum is consistent with a power-law although we find the spectral slope varied between both \xmm\ observations and is also significantly different from that reported previously. Whilst we cannot rule-out intrinsic cold absorption to explain the spectral depression, we favour a dust-free warm absorber. Cold absorption requires a column density $\sim1.4-1.6\times10^{22}\pcmsq$ whilst a warm absorber could have up to $\sim10^{23}\pcmsq$ and an ionization parameter $\sim10^{2}$. The spectrum of \gb\ shows remarkable parallels with that of the $z=4.4$ blazar \pmn, in which the available evidence is also most consistent with a warm absorber model.
\end{abstract}


\section{Introduction}
High-redshift quasars are important objects in the study of the early evolution of massive black holes and the interaction with their host galaxies. A number of radio-loud quasars have been found to show the spectral characteristics typical of blazars. \gb\, at $z=4.72$ \citep{hook98,fabian97,fabian98,fabian99} displays spectral flattening at soft X-ray energies in \rosat\ \citep{boller00} and \bepposax\ \citep{fabian01_gb} data. Here we confirm the flattening with \xmm\ observations. Soft X-ray flattening has also been reported from \rx\ \citep{yuan00} and \pmn\ \citep{fabian01_pmn} at redshifts of $z=4.28$ and $z=4.4$ respectively. \xmm\ data shows only marginal evidence for flattening in the case of \rx\ \citep{grupe04}. Conversely, the effect has been confirmed in \pmn\ \citep{worsley04}. Spectral flattening is consistent with a trend seen in a number of studies of lower-redshift radio-loud quasars \citep[e.g.][]{cappi97,fiore98,reeves00}. 

The most consistent explanation for the flattening is intrinsic absorption with column densities of the order $10^{22-23}\pcmsq$. Intergalactic absorption systems remain an unlikely possibility given their low metallicities. Furthermore, in order to account for the flattening the line of sight to the blazar would need to intercept two or more very high column density systems of which the probability is very low. Certainly this explanation cannot account for the spectral depressions seen in an increasing number of objects. An alternative explanation for flattening is a break in the underlying blazar continuum. Hard X-ray production in blazars is likely to be due to the inverse Compton scattering of seed photons by relativistic electrons in the jet. A spectral break can arise through a low energy cut-off in the electron population or by a peaked seed photon spectrum, however the break appears much too sharp for this to be plausible.

Given an intrinsic absorber hypothesis there remains much discussion about the nature of the absorber. X-ray data alone have been unsuccessful in breaking the degeneracy between a cold (neutral) absorber or one that is warm (ionized). Optical observations however, can play a crucial role in this regard. A number of the blazars showing soft X-ray flattening have optical spectra (corresponding to rest frame UV emission) which indicate very little dust along the line of sight or show any evidence of a Lyman-limit system (for the optical data for \gb\ and \pmn\ see \citealp{hook98} and \citealp{peroux01}). These point to a dust-free yet metal-enriched absorber where hydrogen has been ionized, consistent with the warm absorber such as that seen in Seyfert galaxies \citep[e.g.][]{reynolds97,crenshaw99}. Strong intrinsic C\thinspace\textsc{iv} absorption is common in these galaxies and it is also a feature of \pmn.

\section{Previous observations}

\gb\ was discovered by \citet{hook98} in an optical survey for high-redshift quasars. An X-ray follow-up was conducted by \citet{fabian97} with the \rosat\ HRI and \asca\ \citep{fabian98}. Spectral fits were consistent with a power-law continuum with a photon index $\Gamma=1.29$ although the \asca\ SIS data alone were suggestive of some excess absorption. Subsequent \rosat\ observations showed that X-ray flux varied by a factor of two over a time-scale of around 2.5 days in the rest frame of the source. Radio data also shows variability, with amplitudes of $\sim40/15$ per cent on rest frame time-scales of weeks/days.

More evidence of X-ray variability was seen in a \rosat\ PSPC observation \citep{boller00}, where changes of $\sim25$ per cent occur on rest frame timescales of 1.8 hours. The low-energy sensitivity of the PSPC detector clearly revealed a spectral depression in the soft X-ray band, consistent with an absorbing column density of $N_{\textrm{H}}\sim1.5\times10^{22}\pcmsq$ (if intrinsic to the source). 

Spectral flattening was further confirmed by \bepposax\ \citep{fabian01_gb}. The fitted spectrum was that of a $\Gamma=1.45$ power-law with an intrinsic absorption column density of $N_{\textrm{H}}\sim8\times10^{22}\pcmsq$. Warm absorption models were considered, with column densities of order $N_{\textrm{H}}\sim10^{23.3}\pcmsq$ and ionization parameters of order $\xi\sim10^{2}$ being marginally consistent with results from \bepposax\ as well as those from the previous \rosat\ and \asca\ observations. 


\begin{table}
\centering
\caption{Details of \xmm\ observations.}
\label{exposures}
\begin{tabular}{ccccc}
\hline
Revolution & Date & \multicolumn{3}{c}{Good Exposure Times (s)} \\
&& PN & MOS-1 & MOS-2 \\
\hline
549 & 2002 Dec 09 & 2625  & 4055  & 4260      \\
569 & 2003 Jan 17 & 11534 & 14241 & 14235     \\
\hline
\end{tabular}
\end{table}


\section{\xmmtitle\ observations}

\subsection{Data reduction}

\xmm\ observed \gb\ on 2002 December 9 and again on 2003 January 17 (Table~\ref{exposures}). All observations were performed in full frame imaging mode with the thin filter. The EPIC data reduction was performed with the Scientific Analysis Software (\textsc{sas}) \textsc{v}5.4.1. Periods of background flaring were excluded and the data were filtered in the standard way to include only the recommended events for spectral analysis \citep{ehle03} i.e. using the event pattern ranges $0-4$ and $0-12$ plus the `FLAG=0' and `\#XMMEA\_EM' filters for the PN and MOS respectively. 

Spectra were extracted from circular regions centred on the source of approximate radius $40''$ and $60''$ for the PN and MOS respectively. Backgrounds were extracted from nearby source free regions following the guidelines in \citet{ehle03}. \textsc{SAS} tasks \textsc{arfgen} and \textsc{rmfgen} were used to generate the appropriate response files. Background subtracted spectra were produced and grouped to a minimum of 20 counts in each bin. Spectral analysis was performed using \textsc{xspec} \textsc{v}11.3.0 \citep{arnaud96}. 

\subsection{Spectral properties}

Spectral fits to a simple power-law inclusive of neutral galactic absorption ($N_{\textrm{H}}=1.40\times10^{20}\pcmsq$; \citealp{elvis94}) were made over the energy range $1-12\kev$. The PN and MOS data were fitted simultaneously, but the revolution 549 and 569 observations were treated separately due to source variability. $1-12\kev$ power-law photon indices of $\Gamma=1.88\pm0.05$ and $\Gamma=1.73\pm0.03$ were obtained, with $2-10\kev$ fluxes of $1.22\times10^{-12}\ergpcmsqps$ and $1.09\times10^{-12}\ergpcmsqps$ respectively. Reduced chi-squared values for the fits were $\chi^{2}_{\nu}=0.819$ and $\chi^{2}_{\nu}=0.883$ (with $\nu=100$ and $\nu=283$ degrees of freedom) respectively. These can be compared to only $\chi^{2}_{\nu}=1.06$ ($\nu=385$) with an attempt to fit the same model to the data from both observations simultaneously. 

\begin{figure}
\rotatebox{270}{
\resizebox{!}{\columnwidth}
{\includegraphics{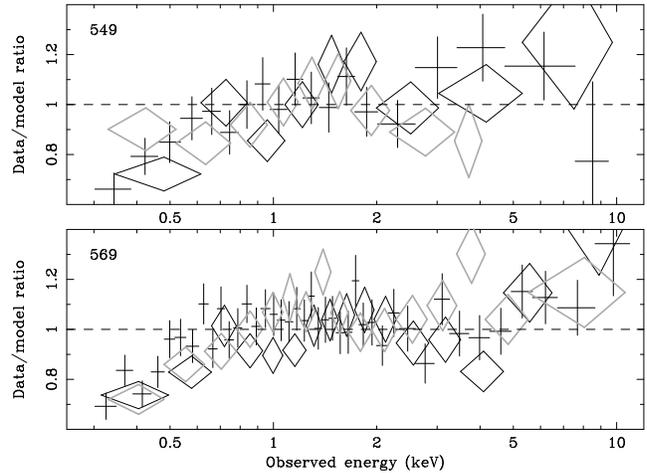}}}
\caption{The upper panel shows the ratio of the revolution 549 data to a simple $\Gamma=1.88$ power-law (including Galactic absorption of $N_{\textrm{H}}=1.40\times10^{20}\pcmsq$; fitted over $1-12\kev$). The lower panel shows the revolution 569 data to a $\Gamma=1.73$ power-law (with the same Galactic correction and fit range). The PN data are shown in both instances as black crosses with the MOS-1/MOS-2 data as black/grey diamonds respectively.}
\label{bothratio}
\end{figure}

A significant depression in the soft X-ray spectrum is seen when the whole $0.3-12\kev$ energy range is considered. The chi-squared statistic of the power-law models rises to $\chi^{2}_{\nu}=1.245$ and $\chi^{2}_{\nu}=1.302$ (for $\nu=173$ and $\nu=430$) respectively. Fig.~\ref{bothratio} shows the ratio of the revolution 549 and 569 data to the $1-12\kev$ fitted power-law. Spectral flattening is clear in both observations, with a break from the power-law occurring at energies $\sim0.6-1\kev$ and the data/model ratio decreasing to $\sim0.7$ at $\lesssim0.4\kev$. 

\begin{figure}
\rotatebox{270}{
\resizebox{!}{\columnwidth}
{\includegraphics{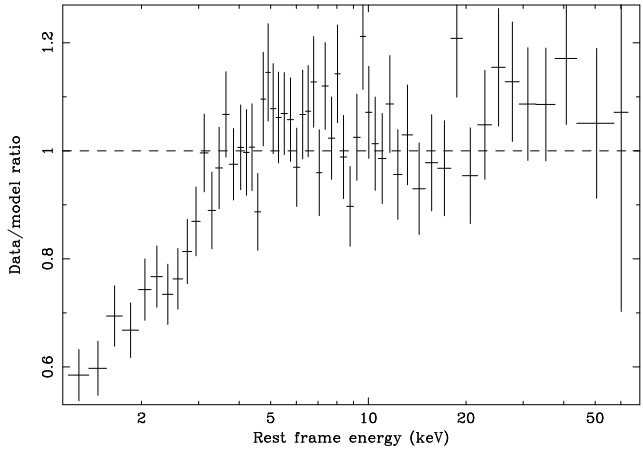}}}
\caption{The combined PN camera data of both observations, plotted as a ratio to a fitted $\Gamma=1.77$ power-law (including Galactic absorption of $N_{\textrm{H}}=1.40\times10^{20}\pcmsq$). The energy has been adjusted to the source rest frame; the observed range is identical to that covered in Fig.~\ref{bothratio}.}
\label{pncomb}
\end{figure}

Even though there has been a significant decrease in flux between the observations (and possibly spectral slope variations) it remains instructive to consider the combined PN spectra in the search for spectral features which may exist at the same energy independently of other variations. To this end the PN camera good events from both observations were merged and a single spectrum extracted. Fig.~\ref{pncomb} shows the ratio of the data to a $1-12\kev$ fitted power-law with $\Gamma=1.77$ (plotted in the rest energy frame of the blazar). Apart from the distinctive spectral flattening there appear to be no statistically significant features. 

\begin{figure}
\rotatebox{270}{
\resizebox{!}{\columnwidth}
{\includegraphics{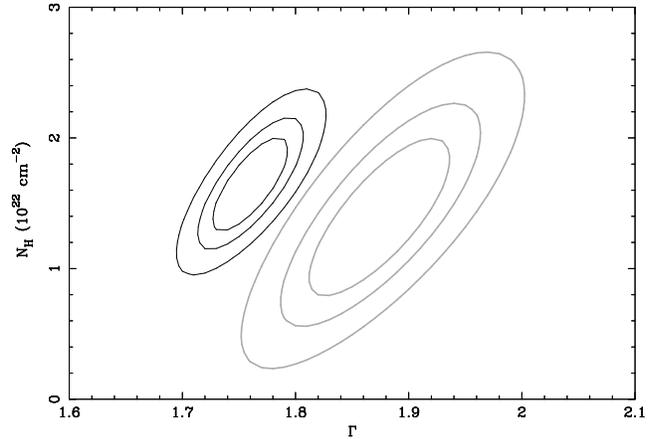}}}
\caption{Confidence contours (68, 90 and 99 per cent) in the photon index and absorber column density for a power-law plus intrinsic cold absorber model (also including galactic absorption). The separate contours for revolutions 549 and 569 are shown in grey and black respectively.}
\label{nhgamma}
\end{figure}

\begin{table*}
\centering
\caption{Summary of the spectral fits performed to the revolution 549 and 549 data. Fits were made in each case, to the PN and MOS spectra jointly over the energy range $0.3-12\kev$. Errors indicated are 1-sigma for one interesting parameter.}
\label{fits}
\begin{tabular}{rccccccc}
\hline
\multicolumn{1}{c}{Model} & local $N_{\rm{H}}$ & intrinsic $N_{\rm{H}}$ & $\Gamma_{1}$ & $E_{\rm{break}}$ & $\Gamma_{2}$ & $\chi^{2}_{\nu}$ & $\nu$ \\
& ($10^{20}\pcmsq$) & ($10^{22}\pcmsq$) && ($\kev$) & & \\
\hline
\multicolumn{8}{l}{Rev. 549 data:}\\
Broken power-law                     & $1.40$ (Galactic) & $0$         & $1.1\pm0.3$ & $0.7\pm0.2$ & $1.83\pm0.03$   & $0.851$ & 171  \\
power-law, free local absorption     & $5.0\pm0.5$       & $0$         &             &             & $1.92\pm0.05$   & $0.840$ & 172  \\
power-law, free intrinsic absorption & $1.40$ (Galactic) & $1.4\pm0.4$ &             &             & $1.87\pm0.04$   & $0.844$ & 172  \\
\hline
\multicolumn{8}{l}{Rev. 569 data:}\\
Broken power-law                     & $1.40$ (Galactic) & $0$         & $0.7\pm0.2$ & $0.6\pm0.1$ & $1.71\pm0.02$   & $0.932$ & 428  \\
power-law, free local absorption     & $5.1\pm0.5$       & $0$         &             &             & $1.79\pm0.03$   & $0.949$ & 429  \\
power-law, free intrinsic absorption & $1.40$ (Galactic) & $1.6\pm0.2$ &             &             & $1.76\pm0.02$   & $0.931$ & 429  \\
\hline
\end{tabular}
\end{table*}

The spectral break can be modelled in a number of ways. Table~\ref{fits} shows the results of broken power-law and power-law plus absorption models. The best-fitting local absorption column density exceeds the expected galactic level at the 7-sigma level. If the excess is modelled as cold absorption at the source then column densities of $1.4-1.6\times10^{22}\pcmsq$ are required. Fig.~\ref{nhgamma} shows the confidence contours in the photon index and column density (for intrinsic cold absorption) for the two observations.

\subsection{Variability}
\label{variability}

Background-corrected lightcurves were extracted for both observations. The good-time-interval lightcurve from the revolution 549 observation is fully consistent with a constant flux (the PN count-rate is $0.73\ctsps$ corresponding to $1.22\times10^{-12}\ergpcmsqps$). Unfortunately the remaining $\sim60$ per cent of the observation was affected by solar flare activity and reliable background subtraction could not be performed. The revolution 549 observation suffered from no flare periods and $\sim12\ks$ of PN data were extracted, again this is consistent (at the 90 per cent level) with a constant flux (a PN count-rate of $0.54\ctsps$ corresponding to $1.09\times10^{-12}\ergpcmsqps$). The $\sim10$ per cent reduction between the observations corresponds to a variability timescale of $\sim7\thinspace\textrm{days}$ in the rest frame. 

In addition to the decrease in luminosity there are differences in the spectral properties of the source between observations. The revolution 549 data are well fit (in the $1-12\kev$ range) by a simple power-law with a photon index of $\Gamma\sim1.9$ whilst this is hardened to $\Gamma\lesssim1.8$ during the revolution 569 observation. The break energy and soft X-ray photon index (and equivalently, the column densities in the absorption models) are in good agreement between the observations.

There would also seem to be significant variations in the spectral shape between the \xmm\ and earlier observations. The continuum slopes (i.e. after correction for any soft X-ray flattening) reported by \rosat\ and \bepposax\ \citep{boller00,fabian01_gb} were $1.4\pm0.2$ and $1.45\pm0.1$ respectively, considerably harder than the $\sim1.7-1.9$ found with \xmm. The spectral break is characterised by the column density required of an intrinsic cold absorber, the \rosat\ and \bepposax\ values of which are $1.5\pm0.3\times10^{22}\pcmsq$ and $7.8^{+8.7}_{-6.0}\times10^{22}\pcmsq$, consistent with each other and the $\sim1.4-1.6\times10^{22}\pcmsq$ found here. 

There is marginal evidence of curvature in the residuals of the \xmm\ fits to the $1-12\kev$ power-law, with excesses at $\sim1-2\kev$ and $\sim5-10\kev$ and a deficit over $\sim2-5\kev$ (Fig.~\ref{bothratio}). The $\gtrsim5\kev$ range appears to have a harder power-law ($\Gamma\sim1.5$), which is in better agreement with the \rosat\ and \bepposax\ results. A steeper $1-12\kev$ slope and curved residuals in the \xmm\ data could possibly then be the result of a $\sim1-3\kev$ variable excess ($\sim5-20\kev$ rest frame) superimposed upon the soft X-ray break.

To investigate further a difference spectrum was formed by subtracting the revolution 569 spectrum from the revolution 549 one. The difference is well-fitted by a power-law model (inclusive of Galactic absorption and intrinsic cold absorption of $N_{\textrm{H}}=1.5\times10^{22}\pcmsq$) with a spectral slope of $\Gamma=2.3\pm0.3$. A two power-law model (again inclusive of Galactic and intrinsic absorption) is also an acceptable fit to both data sets. The soft power-law component has $\Gamma\sim1.8-2.6$ and the hard power-law has $\Gamma\lesssim1.7$. 

The curved residuals, $1-3\kev$ excess and the success of the two power-law model point to the intriguing possibility that the spectrum may contain a additional component. The steepness ($\Gamma\sim2.3$) of the soft component is consistent with the high-energy cut-off of a second inverse-Compton contribution. Possibilities for such a component are an underlying pile-up in the EC seed photon distribution (see \ref{intrinsic}), an EC component with a different external photon source, SSC emission or a bulk Compton component due to inverse-Compton scattering from a cold electron population.

\section{Discussion}

There are two main possibilities for the origin of the spectral flattening observed in \gb. The first, and our favoured explanation, is that the flattening is due to excess intrinsic absorption. The second hypothesis attributes the effect to a break in the underlying blazar emission continuum. These two alternatives are now discussed separately.

\subsection{Absorption}

A galactic origin to the excess column density is extremely unlikely; it would require a factor $\sim3.6$ increase in the galactic column density, highly localised in the direction of \gb. Intergalactic absorption is also highly unlikely, the only possibility being a line of sight that passes through two or more very high column density damped Ly-$\alpha$ systems for which the probability is very low \citep{o'flaherty97,zwaan99}. If an absorber is the origin of the X-ray flattening it must therefore be metal-rich and intrinsic to the source. A cold absorber model requires a column density in the region $1.4-1.6\times10^{22}\pcmsq$.


A warm (ionized) absorber is able to reproduce both low UV and high X-ray column densities. Photoionized gas could produce significant photoelectric absorption in the soft X-ray band (mainly due to oxygen and neon species) whilst remaining transparent at UV wavelengths (hydrogen/helium both ionized). Additionally, any dust grains could be destroyed by the intense photon-flux, resulting in low levels of extinction in the UV band.

\begin{figure}
\rotatebox{270}{
\resizebox{!}{\columnwidth}
{\includegraphics{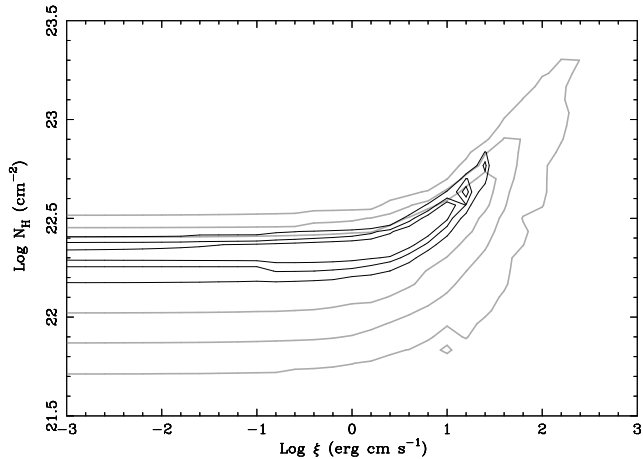}}}
\caption{Confidence contours (68, 90 and 99 per cent) in the photon index and column density for a power-law plus intrinsic cold absorber model (also including galactic absorption). The separate contours for the revolution 549 and 569 observations are shown in grey and black respectively.}
\label{xinh}
\end{figure}

The photoionization code \textsc{cloudy} \citep{ferland98} was used to model the warm absorber hypothesis. Fig.~\ref{xinh} shows the confidence contours in the warm absorber ionization parameter and column density for the two separate observations. Neither set of contours strongly constrains the ionization parameter provided that $\xi\lesssim10^{2}\ergcmps$. The column density is in the range $10^{22}\lesssim N_{\rm{H}}\lesssim10^{23}\pcmsq$. 

It is difficult to break this degeneracy with X-ray observations: The conclusive signature of a warm absorber would be the detection of photoelectric absorption edges, principally those arising from O/Ne. Due to the high-$z$ of the source however, these features are redshifted out of the observed band to energies $\lesssim0.2\kev$. A strong discriminator would be evidence of Lyman-limit absorption at $5217\a$ (the present optical data start at $5970\a$). This would place constraints on the neutral hydrogen column density and thus on the ionization state of the absorber. The lack of an optically-thick Lyman-limit system would immediately place an upper limit to the H\thinspace\textsc{i} column density of $\sim10^{17}\pcmsq$, requiring the warm absorption model to have an ionization parameter $\xi\gtrsim1\ergcmps$.

\subsection{Intrinsic spectral properties}
\label{intrinsic}

The revolution 549 and 569 spectra can be acceptably fitted with a broken power-law model with a break at an energy $\sim0.6-0.7\kev$ down to a flat spectral index $\sim0.7-1.1$. In the rest frame (Fig.~\ref{pncomb}) the break energy is $\sim4\kev$ and is very distinct, occurring over a range of $\sim2\kev$. The $z=4.4$ blazar \pmn\ is similar with a sharp spectral break at a rest frame energy of $\sim4.5\kev$ over an energy range $\lesssim5\kev$. 

The dominant emission components in blazars are beamed synchrotron radiation from the jet (peaking in the infra-red to soft X-ray band) and a high-energy component (dominating the hard X-ray to $\gamma$-ray regimes). The high-energy emission arises through the inverse Comptonization of soft photons by highly relativistic electrons in the jet plasma. Sources of the soft photons are the nuclear optical/UV emission or the local synchrotron radiation within the jet. For the brightest blazars (such as \gb\ and \pmn) the first, known as the `external Compton' (EC) mechanism, is dominant \citep{sikora94}.

A flattening in the spectrum could be expected if there was a low energy cut-off in the energy distribution of the electron population in the jet. In the case of the EC process it is also possible to produce a spectral break by requiring the seed photon distribution to be sharply peaked at and/or rapidly decline below a certain frequency. There are several problems with these explanations: Firstly, there is no evidence of any spectral breaks in the spectra of nearby radio-loud quasars, and secondly, there seems to be no plausible way to account for the sharpness of the break (only a few $\kev$ in the rest frame). For a more detailed discussion of the break-producing mechanisms see e.g. \citet{fabian01_pmn}.

\subsection{Comparisons with other objects}

\begin{figure}
\rotatebox{270}{
\resizebox{!}{\columnwidth}
{\includegraphics{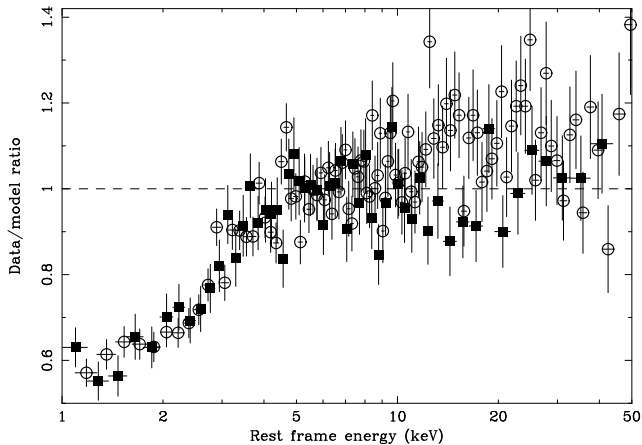}}}
\caption{The ratio of combined PN camera spectra to a hard-band fitted power-law. The filled squares show the data for \gb; the power-law has $\Gamma=1.77$ and was fitted over the observed $1-12\kev$ range. The open circles show the data for \pmn; the power-law has $\Gamma=1.71$ and was fitted over the observed $2-12\kev$ range. The spectra are plotted in their respective rest frame energies. The spectra have also been renormalised to a data/model ratio of unity over the $5-10\kev$ rest frame range.}
\label{compareratio}
\end{figure}

Fig.~\ref{compareratio} shows the rest frame combined PN spectrum of \gb\ (as a ratio to the power-law above the spectral break) along with the equivalent result for \pmn\ \citep{worsley04}. The agreement in the position and structure of the break is remarkable. Since the plots are of ratios to the model the instrumental response (and Galactic absorption) have been removed. In both cases the break from the power-law occurs over a range of $\lesssim5\kev$ in the source frame and to a photon index of $\Gamma\sim1$. Both also show a steeper drop in the flux at $3\kev$. 

If the spectral flattening is indeed due to warm absorption then the absorbers for \gb\ and \pmn\ are almost identical in column density (of $\sim2-4\times10^{22}\pcmsq$). There is also evidence for a similar column in the \xmm\ observations of \rx\ \citep{grupe04} where joint fits to the \xmm\ and previous \asca\ data \citep{yuan00} are consistent with an excess intrinsic column density of $\sim1\times10^{22}\pcmsq$.

\section{Conclusions}

We have studied the X-ray spectrum of the high-redshift, radio-loud blazar \gb, providing definitive confirmation of the soft X-ray spectral flattening as previously observed with \rosat\ and \bepposax. The flattening is not consistent with Galactic or intergalactic absorption and we propose that the effect is due to an absorber that is intrinsic to the source, or to an underlying break in the blazar continuum. 

A continuum break is highly speculative given the present lack of knowledge concerning emission processes in the nuclear regions. Continuum breaks are not seen in nearby radio-loud objects where they should be readily detectable and there is no satisfactory way to account for the sharpness of the break seen. An intrinsic absorber is therefore our favoured explanation. Cold absorption requires a column density $N_{\textrm{H}}\sim1.4-1.6\times10^{22}\pcmsq$, whilst a warm absorber could have a column density of up to $\sim10^{23}\pcmsq$ and an ionization parameter $\xi\sim10^{2}$. X-ray data alone are unable to distinguish between a warm or cold absorber model. Good constraints are expected from future optical observations which could place strong limits on the neutral hydrogen column density. 

It is interesting to note the parallels with the blazar \pmn\ \citep{worsley04} i.e. a relatively blue UV continuum with strong Ly-$\alpha$ and C\thinspace\textsc{iv} emission lines as well as the flat hard X-ray spectral slope and the remarkably similar form of the soft X-ray depression. The \pmn\ X-ray data are consistent with both warm/cold absorption but optical observations rule out a cold absorber scenario. The lack of significant reddening in the UV spectra of both objects is consistent with a dust-free warm absorber. 

Comparisons can also be drawn to the $z=3.27$ radio-loud quasar \pks\ which also shows definite soft X-ray absorption \citep{ferrero03,fiore03}. Both warm and cold absorbers are consistent although ionized absorption is again most able to explain the lack of significant UV reddening. The implied iron abundance of a warm absorber however is at odds with the super-solar abundances implied by broad emission lines. 

The spectrum of \gb\ are suggestive of an additional component contributing to an excess of emission in the $\sim5-10\kev$ rest-frame energy range. There are a number of possibilities for the emission including a pile-up in the electron energy distribution or a feature caused by bulk Comptonization, however the detection is marginal and further work is required. 

\section{Acknowledgments}
Based on observations with \xmm, an ESA science mission with instruments and contributions directly funded by ESA Member States and the USA (NASA). MAW acknowledges support from PPARC. ACF thanks the Royal Society for support. AC thanks the Italian MIUR and INAF for financial support.

\bibliographystyle{mnras} 
\bibliography{mn-jour,worsley_29mar04}

\end{document}